\documentclass[10pt]{article}

\usepackage{amsmath, latexsym, amsfonts, amssymb, amsthm}
\usepackage{graphicx, color,hyperref,dsfont,epsfig,caption,wrapfig,subfig}
\usepackage{pstricks}
\usepackage{datetime}
\usepackage{movie15}
\hypersetup{
    linktoc=page,
    linkcolor=red,          
    citecolor=blue,        
    filecolor=blue,      
    urlcolor=cyan,
    colorlinks=true           
}

\numberwithin{equation}{section}

\newtheorem{theorem}{Theorem}[section]

\theoremstyle{definition}

\renewcommand{\tilde}{\widetilde}          
\DeclareMathSymbol{\leqslant}{\mathalpha}{AMSa}{"36} 
\DeclareMathSymbol{\geqslant}{\mathalpha}{AMSa}{"3E} 
\DeclareMathSymbol{\eset}{\mathalpha}{AMSb}{"3F}     
\renewcommand{\leq}{\;\leqslant\;}                   
\renewcommand{\geq}{\;\geqslant\;}                   

\newcommand{\C}{\mathbb{C}}

\newcommand{\R}{\mathbb{R}}
\newcommand{\Z}{\mathbb{Z}}

\newcommand{\Q}{\mathbb{Q}}

\newcommand{\E}{\mathds{E}}
\renewcommand{\P}{\mathds{P}}

\newcommand{\hf}{\frac{_1}{^2}}

\usepackage{xparse}

\def\bi{\begin{itemize}}
\def\ei{\end{itemize}}

\def\bnum{\begin{enumerate}}
\def\enum{\end{enumerate}}
\def\<#1{\langle #1 \rangle}

\newcommand{\caF}{{\mathcal F}}

\newcommand{\caO}{{\mathcal O}}

\newcommand{\caT}{{\mathcal T}}

\def\ak#1{\textcolor{red}{(Note: {#1})}}

\title{The DOZZ Formula from the Path Integral}

\author{ Antti Kupiainen \footnote{University of Helsinki, Department of Mathematics and Statistics, P.O.
Finland. Supported by the Academy of Finland and ERC Advanced Grant 741487} , R\'emi Rhodes \footnote{Universit{\'e} Paris-Est Marne la Vall\'ee, LAMA, Champs sur Marne, France. Partially supported by grant    ANR-15-CE40-0013 Liouville.} , 
 Vincent Vargas \footnote{ENS Ulm, DMA, 45 rue d'Ulm,  75005 Paris, France. Partially supported by grant    ANR-15-CE40-0013 Liouville.} }

\begin{document}

\maketitle
 
 \begin{abstract}
 We present a rigorous proof of the Dorn, Otto, Zamolodchikov, Zamolodchikov formula (the DOZZ formula) for the 3 point structure constants of Liouville Conformal Field Theory (LCFT) starting from a rigorous probabilistic construction of the functional integral defining LCFT given earlier by the authors and David. A crucial ingredient in our argument is a  
  probabilistic derivation of the reflection relation in LCFT based on a refined tail analysis of Gaussian multiplicative chaos measures. 
   \end{abstract}


\begin{center}
\end{center}
\footnotesize



\normalsize

\section{Introduction}

One of the simplest and at the same time most intriguing  Conformal Field Theories (CFT hereafter) is Liouville CFT  (LCFT hereafter). It first appeared in Polyakov's path integral formulation of String Theory \cite{Pol} and then  in the work of Knizhnik, Polyakov and Zamoldchickov \cite {KPZ} on the relations between CFT's in a fixed background metric and in a  random metric (2d gravity). 
They argued that the correlation functions of a CFT coupled to a random metric are given as products of ordinary CFT correlations and LCFT correlations.

Unlike most CFT's LCFT has an explicit functional integral formulation. More precisely, LCFT is a theory of a scalar field $\phi(z)$ defined on the  Riemann Sphere  $z\in\hat\C= \C \cup \lbrace \infty \rbrace$ equipped with a fixed ``background" metric $g(z)|dz|^2$. The functional integral corresponds to the measure $e^{-S(\phi)}D\phi $ where the Liouville action $S$ is defined by
\begin{equation}\label{actionLiouville}
S(\phi)= \frac{1}{\pi}\int_{\C}|\partial_z\phi |^2dz+  \frac{1}{4\pi}\int_{\C}\big(QR_{{g}} \phi +4\pi \mu e^{2b \phi }\big)\,g(z)d^2z   .
\end{equation}
where $\partial_z=\frac{1}{2}(\partial_x-i\partial_y)$ for $z=x+iy$ (and $\partial_{\bar z}=\frac{1}{2}(\partial_x+i\partial_y)$), $R_g=-4g^{-1}\partial_z\partial_{\bar z}\ln g$ is the scalar curvature  and $Q=b+\frac{1}{b}$. The parameter $b$ is a priori an arbitrary complex number. However, for the probabilistic models  coming from 2d gravity $b\in (0,1)$ which we assume in the sequel\footnote{A rigorous probabilistic construction of \eqref{actionLiouville} is still lacking for general complex $b$.}. Finally $\mu>0$ is the cosmological constant. The primary fields in LCFT are the vertex operators 
$V_\alpha(z):=e^{2\alpha\phi(z)}$ 
and their  correlation functions are formally given for distinct $z_1, \ldots, z_n \in \hat\C$ by
\begin{equation}\label{correLiouvilleintro}
\langle \prod_{i=1}^n V_{\alpha_i}(z_i)\rangle =Z\int  \prod_{i=1}^n 
e^{2\alpha_i\phi(z_i)}
e^{-S(\phi)}D\phi 
\end{equation}
where $\int$ denotes (formal) integration over all scalar fields $\phi$ and $Z$ is an overall normalization to be fixed. The above definition is formal and the rigorous 
construction of this functional integral given by the authors and David \cite{DKRV}, requires a regularization and renormalization procedure discussed in 
Section \ref{thermal}.   

Decisive progress in LCFT came in the 90's as Dorn and Otto \cite{Do} (see also \cite{Do0}) and Zamolodchikov and Zamolodchikov \cite{ZZ} produced an explicit formula for the Liouville three point functions $\langle V_{\alpha_1}(0)  V_{\alpha_2}(1)  V_{\alpha_3}(\infty)  \rangle$, the celebrated DOZZ formula. The DOZZ formula is a rather far reaching and intriguing expression involving special functions from number theory: it is stated in expression \eqref{DOZZformula} below. From this formula and an assumption on the spectrum of the theory, one can construct the higher order correlations of LCFT by the conformal bootstrap procedure of  Belavin-Polyakov-Zamolodchikov (BPZ) \cite{BPZ}; these higher order correlations are expressed in terms of the DOZZ formula and the universal conformal blocks of CFT. More recently, the Liouville three point functions and the conformal blocks were shown  to have a deep relation to four dimensional Yang-Mills theories \cite{AGT}. 

However the exact results for Liouville correlations are not derived from the functional integral but rather from general principles of CFT (BPZ equations, crossing symmetry) coupled to assumptions about the spectrum of LCFT  \cite{Teschnerreview,Rib}. We would also like to signal out the formal derivation of the DOZZ formula  in \cite{Teschnerreview,BytskoTeschner,teschnerproof} from a free field representation of the Liouville vertex operators.  Attempts to use the functional integral were made hard by the fact that the cosmological constant $\mu$ is not a perturbative parameter in the theory as it can be scaled at will: more precisely two different values of $\mu$ produce exactly the same theory\footnote{As long as $\mu>0$; the case $\mu=0$ gives a different theory, the Gaussian Free Field theory.}. Compatibility of the DOZZ formula  with   the semiclassical limit of the functional integral was observed by the Zamolodchikov brothers \cite{ZZ} (see also the recent paper by Harlow-Maltz-Witten \cite{witten} for a more general study of the semiclassical limit) but the general case has resisted attempts.

The purpose of this paper is to sketch a recent mathematical proof \cite{KRV,KRV2017}\footnote{Our proof 
is based on 
\cite{KRV} where we established the BPZ equations for the degenerate field insertions and on 
 \cite{KRV2017}, where we used them along with a probabilistic identification of the reflection principle to prove the DOZZ formula.} 
 of the DOZZ formula that is based on a rigorous probabilistic construction of the LCFT functional integral in 
 \cite{DKRV}. 
The proof involves many technical subtleties on the mathematical side but we believe that its basic structure should be of interest to physicists as it provides a novel approach to LCFT and DOZZ.  The crux of our proof is    a probabilistic understanding of  the reflection symmetry of LCFT. This symmetry which identifies seemingly different primary fields to each other has been one  of the most mysterious  properties of   LCFT emerging from the DOZZ formula.  
 Our proof settles also another controversial issue for LCFT namely the duality property  that the DOZZ formula satisfies with respect to the following two substitutions:
  \begin{equation}\label{dualityrelation}
 b \leftrightarrow \frac{1}{b}, \quad  \mu  \leftrightarrow \tilde{\mu}= \frac{(\mu \pi \ell(b^2)  )^{\frac{1}{b^2} }}{ \pi \ell(\frac{1}{b^2})}. 
\end{equation}
This symmetry of the DOZZ formula is not present in the action \eqref{actionLiouville} defining the theory. Our proof of the DOZZ formula is based on a probabilistic construction of the action \eqref{actionLiouville} and in particular this demonstrates that it is not necessary to add any  dual potential $\tilde\mu e^{\frac{2}{b}\phi}$ to the action in order to make the path integral compatible with this symmetry.

\section{Probabilistic formulation of LCFT}\label{thermal}
In this section we recall the precise formulation of the  Liouville functional integral  as given in  \cite{DKRV}\footnote{In  the literature on LCFT two conventions for the vertex operators are used and in 
\cite{DKRV,KRV,KRV2017} we used the one where the
interaction term is $4\pi \mu e^{\gamma \phi }$ and vertex operators are denoted by $e^{\alpha\phi(z)}$. In the present work we stick to the more common physics conventions. The translation between the two is obtained by repalcing $\gamma$ by $2b$ and $\alpha$ by $2\alpha$. 
}. The functional integral is defined rigorously as a limit of a regularized and renormalized expression where the   quadratic part of the action is defined through  the Gaussian Free Field (GFF hereafter).  It will be convenient to work in  the background  metric $g(z)=\max(|z|,1)^{-4}$ which has the curvature concentrated on the equator $R_g=4 \delta_{|z|=1}$.\footnote{This is no loss since in \cite{DKRV} it was proved that the usual Weyl anomaly formula holds for the variation of the background metric.} We set
 $\phi=c+\varphi$ where we separate the constant zero mode $c$ and let $\varphi$ be orthogonal to the constants in the sense $\int_{\C}\varphi  R_g g d^2z=0$. The field $\varphi$ is then taken Gaussian with covariance (2-point function)
  \begin{equation}\label{hatGformula}
\E [ \varphi(z)\varphi (z')] =\ln\frac{1}{|z-z'|}-\frac{1}{4}(\ln g(z)+\ln g(z')) :=G(z,z')
\end{equation}
where we adopt the standard probabilistic notation $\E[\cdot]$ for average with respect to the randomness of the field $\varphi$  and reserve the notation $\langle\cdot\rangle$ for the Liouville expectation below.  
Define then the regularized field with UV cutoff $\epsilon$
\begin{equation}\label{circleaverage}
\varphi_{\epsilon}(z):=\frac{1}{2\pi i}\oint_{|w|=\epsilon} \varphi(z+w)\frac{dw}{w}
\end{equation}
 and the regularized vertex operators
 \begin{equation}\label{Vdefi}
V_{\alpha, \epsilon}(z)= 
e^{2\alpha c} e^{2\alpha \varphi_{\epsilon}(z)-2{\alpha^2} \E[ \varphi_{\epsilon}(z)^2] }g(z)^{\Delta_\alpha}
\end{equation}
 where $\Delta_\alpha=\alpha(Q-\alpha)$ is the conformal weight. Then the precise definition of the n-point function \eqref{correLiouvilleintro}  for distinct $z_1, \ldots, z_n \in \C$ is
 \begin{equation}\label{correLiouville}
\langle \prod_{i=1}^n V_{\alpha_i}(z_i)\rangle =\lim_{\epsilon\to 0}2\int_{-\infty}^\infty dce^{-2Qc} \E \left [ \prod_{i=1}^nV_{\alpha_i, \epsilon}(z_i)e^{-\mu \int_{\C} V_{b, \epsilon}(z)d^2z } \right ] .
\end{equation}
It was proven in \cite{DKRV} that 
the limit as $\epsilon\to 0$ exists provided $\sum_{i=1}^n\alpha_i>Q$. This condition allows one to integrate over the constant mode \cite{GL} and we arrive at
\begin{equation}\label{correLiouville1}
\langle \prod_{i=1}^n V_{\alpha_i}(z_i)
\rangle =
\mu^{-s} b^{-1}\Gamma(s)\E\, \left [ \frac{\prod_i e^{2\alpha_i\varphi(z_i)-2{\alpha_i^2} \E\, \varphi(z)^2}g(z_i)^{\Delta_{\alpha_i}}}{(\int_{\C} e^{2b\varphi(z)-2{b^2} \E\, \varphi(z)^2 }g(z)d^2z) ^{s} }  \right ]
\end{equation}
where $s=\frac{\sum_{i=1}^n \alpha_i-Q}{b}$ (In the sequel, we will use this convention for $s$ and it should be clear from the context what $s$ refers to.). 
The final step consists of getting rid of the vertex operators in the numerator by  making a shift in the Gaussian field $\varphi$ (called Girsanov theorem in probability and equivalently ``complete the square trick" in statistical physics) 
\begin{equation}\label{shift}
\varphi(z)\to\varphi(z)+2\sum_{i=1}^n\alpha_iG(z,z_i).
\end{equation}
The result is 
\begin{equation}
\langle \prod_{i=1}^n  V_{\alpha_i}(z_i)\rangle  =b^{-1}
\mu^{-s} \Gamma(s)\prod_{j < k} \frac{1}{|z_j-z_k|^{4\alpha_j \alpha_k}}\E\, 
\left[Z(\alpha,{\bf z})^{-s}\right]\label{Z1}
\end{equation}
where 
\begin{equation}\label{Z1a}
Z(\alpha,{\bf z})=
\int_{\C} \prod_{i=1}^ng(z)^{-b\alpha_i}|z-z_i|^{-4b \alpha_i}  dM(z)
\end{equation}
and
\begin{equation}\label{Z1aa}
 dM(z)=\lim_{\epsilon\to 0}e^{2b\varphi_\epsilon(z)-2b^2\E\, \varphi_\epsilon(z)^2}g(z)d^2z.
\end{equation}
Expression \eqref{Z1}, which expresses the LCFT correlations as an expectation with respect to the GFF, is well defined from the point of view of probability theory and is the starting point of our probabilistic study of LCFT (provided the $\alpha_i$ satisfy appropriate bounds). As a matter of fact, expression  \eqref{Z1} was introduced in the mathematical work \cite{DKRV} and is new even with respect to the physics literature\footnote{Building on our work with David \cite{DKRV} and starting with expression \eqref{Z1}, the work of Cao-Rosso-Santachiara-Le Doussal \cite{santa} establishes a link between LCFT correlations for special values of $\alpha_1, \cdots, \alpha_n$ and GMC measures normalized to have mass $1$, i.e. certain Gibbs measures.}. 

The (random) measure $M$ is a much studied object in probability\footnote{We use the standard mathematical terminology measure in the sequel; in statistical physics, one usually  calls a measure a volume form. The measure $M$ is random since the GFF is a field $(\varphi(z,\omega))_{z \in \C}$ where $\omega$ belongs to a probability space.}. The (weak) limit in \eqref{Z1aa} exists almost surely and the limiting measure is called  Gaussian Multiplicative Chaos (GMC hereafter) following its mathematical introduction by Kahane \cite{cf:Kah}. $M$ is singular with respect to the usual volume measure $d^2z$ and it has a nontrivial multifractal spectrum. One interesting consequence of its fractal properties is  that  the function $ \frac{ 1}{|z-z_i|^{4b \alpha_i}}$ is integrable with respect to $M$ around the singularity at $z_i$    if and only if $\alpha_i<\frac{Q}{2}$ \cite{DKRV}. Hence provided $\sum_{i=1}^n\alpha_i>Q$
\begin{equation*}
\langle \prod_{i=1}^n  V_{\alpha_i}(z_i)\rangle \neq 0 \ \ \ {\rm if\ and \ only \ if}\ \ \forall i, \: \alpha_i<\frac{Q}{2}.
\end{equation*}
The bounds 
\begin{equation}\label{Seibergbounds}
\sum_{i=1}^n\alpha_i>Q, \quad \forall i, \: \alpha_i<\frac{Q}{2}
\end{equation}
  are called the {\bf Seiberg} bounds \cite{seiberg}. Note in particular that they imply that $n\geq 3$. Hence LCFT correlation functions are finite only starting with the three point function and then they are non-zero only if all the weights satisfy $\alpha_i<\frac{Q}{2}$. Actually, starting with the formula \eqref{Z1} we prove that the standard Seiberg bounds are not necessary: the threshold $\sum_{i=1}^n \alpha_i=Q$ produces a trivial singularity in the $\Gamma$ function but the expectation $\E[  Z(\alpha,{\bf z})^{-s} ]$ in \eqref{Z1}  is well defined even for some positive exponent $-s$, namely  provided
\begin{equation}\label{extseib}
Q-\sum_{i=1}^n\alpha_i<\min \left ( \frac{1}{b} , \min_{1 \leq i \leq n}(Q-2\alpha_i) \right ).
\end{equation}
This region was  also identified recently in the path integral study of \cite{witten} where it is called region II\footnote{In fact, the authors consider also complex $\alpha_i$.}; in this paper, we will call conditions \eqref{extseib} the extended Seiberg bounds. These conditions can also be seen as quantum analogues of the conditions discovered by Troyanov \cite{Troy} for the existence of smooth metrics with negative curvature with conical singularities at the points $z_i$.

  In \cite{DKRV} it was proven that the expression \eqref{Z1} transforms under M\"obius transformations as a conformal tensor with conformal weights $(\Delta_{\alpha_i}, \Delta_{\alpha_i})$.
 The  three-point function is then determined  up to a constant
\begin{align}
  \langle      \prod_{i=1}^3 V_{\alpha_i}(z_i)   \rangle 
 & =  |z_1-z_2|^{ 2 \Delta_{12}}  |z_2-z_3|^{ 2 \Delta_{23}} |z_1-z_3|^{ 2 \Delta_{13}}C(\alpha_1,\alpha_2,\alpha_3) \label{confinv3}
\end{align}
where we denoted $\Delta_{12}=\Delta_{\alpha_3}-\Delta_{\alpha_1}-\Delta_{\alpha_2}$ etc... 
The three point structure constants $C(\alpha_1,\alpha_2,\alpha_3)$\footnote{The structure constants are also sometimes denoted $\langle V_{\alpha_1}(0)  V_{\alpha_2}(1)  V_{\alpha_3}(\infty)  \rangle$ as justified by the limit \eqref{Climit} below.} are a fundamental object in LCFT. We can write them in terms of the GMC by noting first that  
\begin{align}
C(\alpha_1,\alpha_2,\alpha_3)&=\lim_{z_3\to\infty} |z_3|^{4 \Delta_3} \langle    V_{\alpha_1}(0) V_{\alpha_2}(1)V_{\alpha_3}(z_3) \rangle \label{Climit}
\end{align}
and then using \eqref{Z1} we get:
\begin{equation}\label{cdefi}
C(\alpha_1,\alpha_2,\alpha_3)=b^{-1}\mu^{-s} 
\Gamma(s)  \E\, [ \rho(\alpha_1,\alpha_2,\alpha_3)^{-s} ]
\end{equation}
where recall that  $s=\frac{\bar\alpha 
-Q}{b}$  (set $\bar\alpha=\sum_{i=1}^3\alpha_i$) and
\begin{equation}\label{cdefi1}
\rho(\alpha_1,\alpha_2,\alpha_3)= 
 \int_{\C}  \frac{1}{ |z|^{4b \alpha_1}  |z-1|^{4b \alpha_2}  } g(z)^{-b\bar\alpha}dM(z).
\end{equation}
M\"obius invariance  fixes the four point function up to a single function depending on the cross ratio of the points. We obtain 
\begin{align}
 \langle   
  \prod_{i=0}^3 V_{\alpha_i}(z_i)  \rangle 
 & = |z_3-z_0|^{- 4 \Delta_{0}}  |z_2-z_1|^{ 2 (\Delta_3-\Delta_2-\Delta_1-\Delta_{0})  } |z_3-z_1|^{2(\Delta_2+\Delta_{0} -\Delta_3 -\Delta_1  )} \\ &\times |z_3-z_2|^{2 (\Delta_1+\Delta_{0}-\Delta_3-\Delta_2)} G\left ( \frac{(z_0-z_1)(z_2-z_3)}{ (z_0-z_3) (z_2-z_1)}  \right )  \label{confinv}
\end{align}
where the labeling of points is for later convenience.
We can recover  the function $G$ as the following limit
\begin{align}
G(z)&= 
\lim_{ z_3 \to\infty }|z_3|^{4 \Delta_3} \langle    V_{\alpha_0}(z)   V_{\alpha_1}(0) V_{\alpha_2}(1)V_{\alpha_3}(z_3)  \rangle   \label{Glimit}.
\end{align}
Using \eqref{Z1} this becomes\footnote{We stress the ${\alpha_0} $ dependence since we will vary it in what follows.}
  $$
  G(z)=|z|^{-4\alpha_0\alpha_1}   |z-1|^{-4\alpha_0\alpha_2}  \mathcal{T}_{\alpha_0}(z)
  $$ 
  where $ \mathcal{T}_{\alpha_0}(z)$ is given by (where following our conventions, $s$ is here given by $s=\frac{\sum_{i=0}^3 \alpha_i-Q}{b}$)
\begin{align}\label{Tdefi}
  \mathcal{T}_{\alpha_0}(z) &=
b^{-1} \mu^{-s} 
 \Gamma(s)  \E \,[ r(z)^{-s}]
 \end{align}
and
\begin{equation}\label{Rdefi}
r(z)= \int_{\C} \frac{1
}{|y-z|^{4b\alpha_0}  |y|^{4b \alpha_1} |y-1|^{4b \alpha_2} }g(z)^{-b\bar\alpha}dM(y).
\end{equation}

\section{BPZ equations}\label{bpzsec}

Starting with the probabilistic expression \eqref{Z1} we proved in \cite{KRV} that the correlations satisfy conformal Ward identities and the  Belavin-Polyakov-Zamolodchikov (BPZ) equations for insertions of degenerate fields. Recall that \eqref{Z1} expresses the correlations as the expectation (average) with respect to some (complicated) functional of the GFF; nevertheless, expression \eqref{Z1} enables to use many tools on Gaussian fields. In particular, the proof of the Ward and BPZ identities are based on standard integration by parts formulas for the GFF $\varphi$. In CFT there are two  level two degenerate fields and in the  LCFT these are given by the vertex operators $V_{\alpha_0}$ where $\alpha_{0}=-\frac{b}{2}$ and  $\alpha_{0}=-\frac{1}{2b}$ respectively.
In \cite{KRV} we proved  the validity of following linear differential equations originally found in \cite{BPZ}:
\begin{align}\label{bpzeq}
 (\frac{1}{4\alpha_{0}^{2}}\partial_{z}^2   + \sum_{k=1}^n \frac{\Delta_{\alpha_k}}{(z-z_k)^2}   +  \sum_{k=1}^n \frac{1}{z-z_k}  \partial_{z_k}  )\langle V_{\alpha_{0}}(z)   \prod_{i=1}^n V_{\alpha_i}(z_i)   \rangle    =  0.
\end{align}
Let us specialize to the case $n=3$. Then \eqref{bpzeq} becomes the hypergeometric equation for the  function $\mathcal{T}_{\alpha_0}$ (defined by \eqref{Tdefi}):
\begin{equation}\label{hypergeo}
z(1-z)\partial_{z}^2 \mathcal{T}_{\alpha_0}(z)+  ({ C}-z({ A}+{ B}+1))\partial_z \mathcal{T}_{\alpha_0}(z) -{ A}{ B} \mathcal{T}_{\alpha_0}(z)=0
\end{equation}
where
\begin{align}\label{defabcfirst}
{ A}&=\alpha_0 (Q-2\bar\alpha)-\hf,  \quad { B}=\alpha_0 (Q-2(\alpha_1+\alpha_2-\alpha_3))+\hf,  \quad C=1+2\alpha_0 (Q-2\alpha_1).
\end{align}
This equation has two holomorphic solutions defined on  $\mathbb{C} \setminus \lbrace (-\infty,0) \cup (1,\infty) \rbrace$:
 \begin{equation}\label{Fpmdef}
F_{-}(z)= {}_2F_1({ A},{  B},{ C},z), \quad F_{+}(z)= z^{1-{ C}} {}_2F_1(1+{ A}-{ C},1+{ B}-{ C},2-{ C},z)
\end{equation}
where $_2F_1(A,B,C,z)$ is given by the standard hypergeometric series (which can be extended holomorphically on $\mathbb{C} \setminus  (1,\infty) $). From the probabilistic representation \eqref{Tdefi}, $ \caT_{\alpha_0}(z)$ is real, single valued and twice differentiable in $\mathbb{C} \setminus \{0,1\}$ and we proved in \cite{KRV}  (Lemma 4.4) that these observations entail that the space of such solutions is one dimensional and given by 
\begin{equation}\label{Tsolution}
\caT_{\alpha_0}(z)= 
\lambda_1 | F_{-}(z) |^2+
\lambda_2| F_{+}(z) |^2
\end{equation}
where  the coefficients satisfy 
\begin{equation}\label{Fundrelation}
\lambda_1/\lambda_2= 
- \frac{\Gamma(C)^2  \Gamma(1-A)  \Gamma(1-B)  \Gamma(A-C+1)  \Gamma(B-C+1) }{ \Gamma(2-C)^2  \Gamma(C-A)  \Gamma(C-B)  \Gamma (A)  \Gamma(B) }
\end{equation}
provided
$C \in \R \setminus \Z$ and $C-A-B \in  \R \setminus \Z$. 

We stress that this result involves no assumptions on crossing symmetry or the like: it follows directly from the probabilistic expression for the four point function and the regularity that we prove using it. But much more can be derived from the formula \eqref{Tdefi}. Indeed, one can find both $\lambda_1$ and  $\lambda_2$ in terms of structure constants by doing an asymptotic analysis of  \eqref{Tdefi} as $z\to 0$. Then the relation \eqref{Fundrelation} will imply nontrivial relations for the structure constants. First,  from \eqref{Tdefi} and  \eqref{cdefi} we obtain that 
\begin{align}\label{Tsolution1}
\lambda_1=\mathcal{T}_{\alpha_0}(0)=C(\alpha_1+\alpha_0,\alpha_2,\alpha_3).
 \end{align}
Consider now the case  $\alpha_{0}=-\frac{b}{2}$. In this case it was shown in \cite{KRV} that one may derive the asymptotics of  $\mathcal{T}_{-\frac{b}{2}}(z)$ as $z\to 0$ so as to determine $\lambda_2$ as well provided $\alpha_1+\frac{b}{2}<\frac{Q}{2}$:
\begin{align}\label{Tsolution2}
\lambda_2=B(\alpha_1)C(\alpha_1+\frac{b}{2},\alpha_2,\alpha_3)
 \end{align}
where
\begin{align}\label{Bdefin}
B(\alpha)= -\mu  \frac{\pi}{  l(b^2) l(2b \alpha)  l(2+b^2- 2b \alpha) }
 \end{align}
 and  we use the notation 
$l(x)=\Gamma(x)/\Gamma(1-x)$\footnote{This function is often denoted $\gamma(x)$ in the standard physics literature.}.
Combining with \eqref{Fundrelation} we get 
\begin{align}\label{3pointconstanteqintro}
C(\alpha_1+\frac{b}{2},\alpha_2,\alpha_3)&=- \frac{_1}{^{\pi \mu}} \mathcal{A}(b)C(\alpha_1-\frac{b}{2},\alpha_2,\alpha_3)
\end{align}
with 
\begin{align}\label{Aformula}
\mathcal{A}(\chi)=\frac{l(-\chi^2)  l(2\chi\alpha_1) l(2\chi\alpha_1-\chi^2 )  l(\chi(\bar{\alpha}-2\alpha_1- \chi) )   }{l( \chi (\bar{\alpha}-\chi - Q)  ) l( \chi (\bar{\alpha}-2\alpha_3-\chi ))  l( \chi (\bar{\alpha}-2\alpha_2-\chi )) }.
\end{align}
Furthermore studying similar asymptotics as $z\to 1$ we derive the relation
 \begin{equation} \label{YetaFundamentalrelation}
 C_\gamma (\alpha_1-\frac{b}{2}, \alpha_2, \alpha_3)= T(\alpha_1,\alpha_2,\alpha_3)  C(\alpha_1,\alpha_2 +\frac{b}{2}, \alpha_3)
\end{equation}
where
$T$ is  given by the following formula 
\begin{equation} \label{defT0}
T(\alpha_1,\alpha_2,\alpha_3)= - \mu \pi \frac{l(A) l(B)}{ l(C) l(A+B-C)l(-b^2) l(2b\alpha_2)  l(2+b^2- 2b\alpha_2)}.
\end{equation}
where $A,B,C$ are given by \eqref{defabcfirst} with $\alpha_0=-\frac{b}{2}$.
\vskip 2mm
The relations  \eqref{3pointconstanteqintro} and \eqref{YetaFundamentalrelation} were originally derived in \cite{teschner} by assuming (i) BPZ equations, (ii) the diagonal form of the solution \eqref{Tsolution} (iii) crossing symmetry . We want to stress that our proof makes no such assumptions, in fact  (i)-(iii) are theorems.
 However they are valid under the condition of the Seiberg bounds i.e.  for  \eqref{3pointconstanteqintro} we need $\alpha_1+\frac{b}{2}<\frac{Q}{2}$. We will later turn to what happens when $\alpha_1+\frac{b}{2}>\frac{Q}{2}$. This is one of the main points of our derivation of the DOZZ formula.

\section{The DOZZ formula and the Reflection Relation}\label{DOZZ}

The functional integral gives an unambiguous definition for vertex operator correlations for real weights $\alpha_i$ satisfying the Seiberg bounds \eqref{Seibergbounds}. However, there are good reasons to believe that they may be analytically continued to a much larger set of weights. 

Consider the three point structure constant  \eqref{cdefi}.
Since the Seiberg bounds require $-s$ to be negative an explicit evaluation of the expectation is not obvious. Based on the fact that for $-s$ a positive integer the expectation may be evaluated by Wick theorem and that the resulting integrals are reminiscent  of the work of Dotsenko and Fateev,  Dorn and Otto \cite{Do} and  Zamolodchikov and Zamolodchikov \cite{ZZ} came up with the following remarkable candidate for the formula for the structure constant, the so called DOZZ formula:
\begin{equation}\label{DOZZformula}
C_{{\rm DOZZ}}(\alpha_1, \alpha_2,\alpha_3)  =\hat\mu^{
-s}   \frac{\Upsilon_b'(0)\Upsilon_b(2\alpha_1) \Upsilon_b(2\alpha_2) \Upsilon_b(2\alpha_3)}{\Upsilon_b({{\alpha_1+\alpha_2+\alpha_3}-Q}) 
\Upsilon_b({{\alpha_1+\alpha_2}-\alpha_3}) \Upsilon_b({{\alpha_2+\alpha_3}-\alpha_1}) \Upsilon_b({\alpha_1+\alpha_3}-\alpha_2)   }
\end{equation}
where recall that in this context $s=\frac{\bar{\alpha}-Q}{b}$ ($\bar{\alpha}= \alpha_1+\alpha_2+\alpha_3$) with $\hat\mu$  defined  by
\begin{equation}\label{hatmu}
\hat\mu  = \pi \mu l(b^2)  b^{2(1 -b^2)}
\end{equation}
(recall that $l(x)=\Gamma(x)/\Gamma(1-x)$) and  $\Upsilon_b$ is an entire analytic function with simple zeros (for $b^2 \not \in \Q$) at the values $-b \mathbb{N}-\frac{_1}{^b} \mathbb{N}$ and $Q+b \mathbb{N}+\frac{_1}{^b} \mathbb{N} $.  It is given by the integral formula
\begin{equation}\label{def:upsilon}
\ln \Upsilon_b(z)  = \int_{0}^\infty  \left ( \Big (\frac{Q}{2}-z \Big )^2  e^{-t}-  \frac{( \sinh( (\frac{_Q}{2}-z )\frac{t}{2}  )   )^2}{\sinh (\frac{t b}{2}) \sinh( \frac{t}{2b} )}    \right ) \frac{dt}{t}.
\end{equation}
The main result of \cite{KRV2017} is:
\begin{theorem} The probabilistic expression $C(\alpha_1,\alpha_2,\alpha_3)$ defined by \eqref{cdefi} for real $\alpha_1,\alpha_2,\alpha_3$ following the extended Seiberg bounds \eqref{extseib} (with $n=3$) satisfies the DOZZ formula \eqref{DOZZformula} and $C_{{\rm DOZZ}}(\alpha_1, ,\alpha_2,\alpha_3)$ is the unique analytic continuation of $C(\alpha_1,\alpha_2,\alpha_3)$ to $\C^3$.
\end{theorem}
An essential role in the proof is an identification in probabilistic terms of  the {\it reflection coefficient} of LCFT. It has been known for a long time that in  LCFT the following reflection relation should hold in some sense:
\begin{equation}\label{rrel}
V_\alpha=R(\alpha) V_{Q-\alpha}
\end{equation} 
where $R(\alpha) $ is a numerical  reflection coefficient. Indeed the DOZZ formula is compatible with the following form of \eqref{rrel}:
\begin{equation}\label{rrel1}
C_{{\rm DOZZ}}(\alpha_1,\alpha_2,\alpha_3)=
R_{{\rm  DOZZ}}(\alpha_1)C_{{\rm {\rm DOZZ}}}(Q-\alpha_1,\alpha_2,\alpha_3) 
\end{equation} 
with 
\begin{equation}\label{defRDOZZ}
R_{{\rm DOZZ}}(\alpha)=- (\pi \: \mu \:  l(b^2)  )^{\frac{ (Q -2\alpha)}{b}} \frac{\Gamma(-b(Q-2\alpha))\Gamma(-\frac{ (Q-2\alpha)}{b})}{\Gamma(b(Q-2\alpha)\Gamma(\frac{(Q-2\alpha)}{b})}.
\end{equation}
The mystery of this relation lies in the fact that the probabilistically defined $C(\alpha_1,\alpha_2,\alpha_3)$ (defined by \eqref{cdefi}) {\it vanish} if any of the  $\alpha_i\geq \frac{Q}{2}$ and $\sum_{i=1}^3 \alpha_i >Q$ (if any of the  $\alpha_i\geq \frac{Q}{2}$ and $\sum_{i=1}^3 \alpha_i < Q$ then \eqref{cdefi} is infinite); more generally, expression \eqref{cdefi} does not satisfy the DOZZ formula if $\alpha_1,\alpha_2,\alpha_3$ do not satisfy condition \eqref{extseib}.  One might think this problem may be fixed by adding an extra renormalization in the regularized vertex operator \eqref{Vdefi} when taking the limit in \eqref{correLiouville}. It is indeed possible to renormalize $e^{2 \alpha\varphi}$ so that the limit is nonzero; for instance when $\alpha=\frac{Q}{2}$ the normal ordered exponential \eqref{Vdefi} has to be multiplied by a $\sqrt{-\ln\epsilon}$ factor \cite{DKRV2}. However, the resulting correlation functions do not satisfy the reflection relation nor the DOZZ formula (similarly, one can renormalize appropriately the vertex operators when $\alpha>\frac{Q}{2}$ but the resulting limit ``freezes" to the same value as the  $\alpha=\frac{Q}{2}$ case).

The clue for how to proceed to access the reflection relation is to go back to expression \eqref{Tsolution} for the degenerate four point function  where the $\lambda_i$ are given by \eqref{Tsolution1} and \eqref{Tsolution2}. These relations are consistent with the fusion rule
\begin{equation}
V_{-\frac{b}{2}}V_{\alpha}=V_{\alpha-\frac{b}{2}} +B(\alpha)V_{\alpha+\frac{b}{2}}\label{fusion}
\end{equation}
 for $\alpha+\frac{b}{2}< \frac{Q}{2}$. How about $\alpha+\frac{b}{2}> \frac{Q}{2}$? Using the DOZZ formula one can check \cite{Rib}  that $B(\alpha)=R_{{\rm DOZZ}}(\alpha)/R_{{\rm  DOZZ}}(\alpha+\frac{b}{2})$ so that combining with \eqref{rrel} the fusion rule \eqref{fusion} becomes
\begin{equation}\label{fusion1}
V_{-\frac{b}{2}} V_\alpha=V_{\alpha-\frac{b}{2}}+R(\alpha)V_{Q-\alpha-\frac{b}{2}}.
\end{equation}  
which now makes sense from the probabilistic perspective if  $\alpha+\frac{b}{2}> \frac{Q}{2}$ and provided we can find a probabilistic expression for $R$ (recall that condition $\alpha+\frac{b}{2}> \frac{Q}{2}$ is equivalent to $Q-\alpha-\frac{b}{2}< \frac{Q}{2}$ and one can define $V_{Q-\alpha-\frac{b}{2}}$ in a probabilistic way). This is indeed what we prove. More precisely we prove the following extension of  \eqref{Tsolution}, \eqref{Tsolution1}, \eqref{Tsolution2}:
\begin{theorem} \label{thetheo4pointexpression11one}
 Let $\chi=\frac{b}{2}$ or $\chi=\frac{1}{2b}$ and $\alpha_1+\chi>\frac{Q}{2}$. Then the solution of the BPZ equation is given by
\begin{equation}\label{theo4pointexpression11one}
\caT_{-
\chi}(z)= C(\alpha_1-
\chi, \alpha_2,\alpha_3)  |F_{-}(z)|^2 +  R(\alpha_1)C(Q-\alpha_1-\chi 
,\alpha_2,\alpha_3) |\caF_{+}(z)|^2
\end{equation}
where the function $R(\alpha)$ has a probabilistic definition given below. 
\end{theorem} 
The proof of this theorem is  a nontrivial asymptotic analysis of the fusion $z\to 0$, which we will explain in  Section \ref{asyana}. There it is shown that $R(\alpha_1)$ is determined by the behaviour of the GMC integral near the singularity $\alpha_1$.  Let $D$ be any neighborhood of the origin and consider the random variable
$$
Z(\alpha):=\int_D|z|^{-4b\alpha}dM(z).
$$ 
Then we prove that 
the tail behaviour of $Z(\alpha)$ is given for $\alpha \in (\frac{b}{2},\frac{Q}{2})$ by
\begin{equation}\label{theo4pointexpression1111}
\P(Z(\alpha)>x)=\bar R(\alpha)x^{-\frac{Q-2\alpha}{b}}+o(x^{-\frac{Q-2\alpha}{b}})
\end{equation}
where $ \bar R(\alpha)$ is independent of $D$. The reflection coefficient is then given for $\alpha \in (\frac{b}{2},\frac{Q}{2})$ by
\begin{equation}\label{deffullR}
R(\alpha)= \mu^{\frac{(Q-2\alpha)}{b}}   \Gamma (-\frac{Q-2\alpha}{b})  \frac{Q-2\alpha}{b} \bar R(\alpha). 
\end{equation}
The tail expansion \eqref{theo4pointexpression1111} is obtained by decomposing the Gaussian field $\varphi$ into a radial and an angular part
\begin{equation*}
\varphi(z)= \varphi_{|z|}(0)+\tilde\varphi(z)
\end{equation*}
where  the radial component $\varphi_{|z|}(0)$ is given by \eqref{circleaverage}) and the angular part $\tilde\varphi(z)$ is independent of the radial one. We show that the tail of the random variable $Z(\alpha)$ is ruled by the large values of the radial component and is given by the tail behaviour of the random variable $\mathcal{I}  e^{2b \mathcal{M}}$ where
$$
\mathcal{M}=\max_{|z|\leq 1} (\varphi_{|z|}(0)+(Q-2\alpha)\ln|z|)$$
and $\mathcal{I}$ is an independent random variable whose tail has a faster decay rate. The law of the radial $|z|\mapsto \varphi_{|z|}(0)$ is that of a Brownian motion in logarithmic time $B_{-\ln|z|}$ as can be seen with a simple check of covariance in \eqref{hatGformula}. The random variable $\mathcal{M}$ is then a standard object in probability theory and it has exponential law with parameter $2(Q-2\alpha)$, producing
$$\P(\mathcal{I}  e^{2b \mathcal{M}}>x)\approx \E[\mathcal{I}^{\frac{Q-2\alpha}{b}}]x^{-\frac{Q-2\alpha}{b}}.$$
$ \bar R(\alpha)$ is then given by $\E[\mathcal{I}^{\frac{Q-2\alpha}{b}}]$ and it has an explicit expression given   in appendix \ref{PRR} in terms of the angular part of the field near $0$.


For later purposes, let us mention that  the same asymptotic analysis of the fusion $z\to 1$ produces a relation similar to \eqref{theo4pointexpression11one} with $\caT_{-\chi}(z)$ expressed in terms of a basis of solutions (hypergeometric series)  around $z=1$. Identifying both relations   yields the crossing relations when $\chi=\frac{1}{2b}$
\begin{align} \label{Yeta2}
 C (\alpha_1-\frac{1}{2b}, \alpha_2, \alpha_3)=& \tilde T(\alpha_1,\alpha_2,\alpha_3) R(\alpha_2) C(\alpha_1,Q-\alpha_2 -\frac{1}{2b}, \alpha_3)\\
 R(\alpha_1) C(Q-\alpha_1 -\frac{1}{2b},\alpha_2, \alpha_3)=&L( \alpha_1, \alpha_2, \alpha_3)R(\alpha_2) C(\alpha_1,Q-\alpha_2 -\frac{1}{2b}, \alpha_3)\label{Yeta3}
\end{align}
where
$\tilde T,L$ are  given by the following formula 
\begin{equation} \label{defT2}
\tilde T(\alpha_1,\alpha_2,\alpha_3)=   \frac{l(\tilde A) l(\tilde B)}{ l(\tilde C) l(\tilde A+\tilde B-\tilde C)},\quad L(\alpha_1,\alpha_2,\alpha_3)=   \frac{l(\tilde C-1) l(\tilde C-\tilde A-\tilde B+1)}{ l(\tilde C-\tilde A) l(  \tilde C-\tilde B)},
\end{equation}
\begin{align}\label{defabc}
\tilde{ A}&=\frac{1}{2b} (2\bar\alpha -Q-\frac{2}{b})-\frac{1}{2},  \quad \tilde{ B}=\frac{1}{2b} (2(\alpha_1+\alpha_2-\alpha_3)-Q)+\frac{1}{2},  \quad \tilde C=1-\frac{1}{b} (Q-2\alpha_1).
\end{align}

\medskip

Let us finally remark that the reflection coefficient has an interesting interpretation in terms of a renormalized two point function. Indeed, we prove that $C(\alpha,\alpha,\epsilon)$ given by  \eqref{cdefi} is defined for $\alpha\in (\frac{b}{2},\frac{Q}{2})$ and $\epsilon>0$ and then
\begin{align}
\lim_{\epsilon\to 0}\epsilon C(\alpha,\alpha, \epsilon) =4R(\alpha).  \label{2pointf}
 \end{align}
 Given \eqref{theo4pointexpression1111} it is not hard to see why \eqref{2pointf} is true. Indeed, from \eqref{cdefi} and \eqref{cdefi1} we have 
 \begin{align*}
 \epsilon C(\alpha,\alpha, \epsilon)\sim_{\epsilon\to 0}2\epsilon  \mu^{\frac{(Q-2\alpha)}{b}}   \Gamma (-\frac{Q-2\alpha}{b}) \E Z(\alpha)^{\frac{Q-2\alpha-\epsilon}{b}}\to4R(\alpha) 
 \end{align*}
where the factor $2$ comes from the two $\alpha$ singularities in the GMC integral. For later purpose let us note that a similar argument gives for $\alpha_1>\alpha_2$:
 \begin{align}
 \epsilon C(\alpha_1,\alpha_2, \alpha_1-\alpha_2+\epsilon)\sim_{\epsilon\to 0}\epsilon  \mu^{\frac{(Q-2\alpha_1)}{b}}   \Gamma (-\frac{Q-2\alpha_1}{b}) \E\, Z(\alpha_1)^{\frac{Q-2\alpha_1-\epsilon}{b}}\to 2R(\alpha_1).  \label{2pointfff}
 \end{align}
 One can check from the DOZZ formula that the relations \eqref{2pointf}, \eqref{2pointfff} hold if we replace $C$ and $R$ by  $C_{{\rm DOZZ}}$ and $R_{{\rm DOZZ}}$.

\section{Solving for $R$}
It was realized by Teschner \cite{teschner} that the periodicity relation  \eqref{3pointconstanteqintro} and its counterpart 
\begin{align}\label{3pointconstanteqintrodual}
C(\alpha_1+\frac{1}{2b},\alpha_2,\alpha_3)&=- \frac{1}{\pi \mu} \mathcal{A}(\frac{1}{b})C(\alpha_1-\frac{1}{2b},\alpha_2,\alpha_3)
\end{align}
coming from the degenerate field $e^{-\frac{1}{b}\phi}$ insertion  determine the structure constants provided the latter are defined for all real weights $\alpha$. Hence we must address the question how to extend them. This will be done by using $R$ to analytically continue  the structure constants to the region violating the extended Seiberg bounds \eqref{extseib}. To do this we first need to prove   the structure constants are analytic in the region allowed by the extended Seiberg bounds \eqref{extseib}.

\subsection{Analyticity of the structure constants}
One of the basic axioms in the bootstrap approach to LCFT \cite{Rib} is analyticity of the correlation functions in the weights $\alpha_i$ as exemplified by the {\rm DOZZ} formula. One might think this is obvious once we have controlled the real weights since at a formal level one has $|e^{2 \alpha\phi}|=e^{2 {\rm Re}\alpha\phi}$. 
However this argument overlooks the renormalization in \eqref{Vdefi}: indeed since $ \E[ \varphi_{\epsilon}(z)^2]=\ln\epsilon^{-1}+\caO(1)$ we see that for complex $\alpha$ one has $|V_{\alpha,\epsilon}|\sim \epsilon^{-2 ({\rm Im}\alpha)^2}V_{{\rm Re}\alpha}$ so that 
 the contribution of the phase cannot be bounded in modulus. Furthermore, if we look at the formula \eqref{Z1} for the correlation function taking $\alpha_i$ complex looks problematic as the integrand in \eqref{Z1a} is no more positive and could have zeros.  We thus need a more subtle analysis of the  renormalization procedure  by means of probabilistic arguments so as to propagate analyticity  of the $\epsilon$-regularized correlations to the limit. Details can be found in \cite{KRV2017} where it is shown that the   Liouville correlations are  analytic in a complex neighbourhood of the real weights $\alpha_i$ satisfying the extended Seiberg bounds \eqref{extseib}. The extension in \cite{KRV2017} to complex values of the $\alpha_i$ is reminiscent of the analysis of the Liouville path integral performed in \cite{witten} at the semiclassical level.

\subsection{Analyticity of the reflection coefficient}
Next we prove that the reflection coefficient  $R(\alpha)$ is analytic in the region  $\alpha\in (\frac{b}{2},\frac{Q}{2})$. This looks more problematic than the analyticity of the  structure constants. Indeed proving analyticity from the probabilistic representation for $R$ \eqref{defunitR} or   from the renormalized two-point function expression  \eqref{2pointf}  seems hard. The key observation for $R$-analycity is the following novel representation of $R$ in terms of the structure constants: for $\alpha\in (\frac{b}{2},\frac{Q}{2})$ we have
\begin{equation}\label{randc}
R(\alpha)=  \mu \pi \frac{l(A) l(B)}{ l(C) l(A+B-C)l(-b^2) l(b^2) 4b }C(\alpha,b, \alpha).
\end{equation}
where $A,B,C$ are given by \eqref{defabcfirst} with $\alpha_0=-\frac{b}{2}, \alpha=\alpha_1=\alpha_3, \alpha_2=\frac{b}{2}$.
This relation  can be derived from the crossing identity \eqref{YetaFundamentalrelation} evaluated for $\alpha_1=\alpha_3=\alpha$
\begin{equation}\label{YetaFundamentalrelationbis}
C (\alpha-\frac{b}{2}, \alpha_2, \alpha)= T(\alpha,\alpha_2,\alpha)  C(\alpha,\alpha_2 +\frac{b}{2}, \alpha).
\end{equation}
The parameters are constrained by the extended Seiberg bounds, yet there is enough room to get close to the region where this relation blows up, namely $\alpha_2\to \frac{b}{2}$, for fixed $\alpha\in (\frac{b}{2},\frac{Q}{2})$. The blow-up of the left hand side is obtained from \eqref{2pointfff}:
$$C (\alpha-\frac{b}{2}, \alpha_2, \alpha)=\frac{2R(\alpha)}{\alpha_2-\frac{b}{2}}+\caO(1)
$$
whereas on the right hand side of \eqref{YetaFundamentalrelationbis}  the blowing up comes from the $T$ function, which is explicit by \eqref{defT0}. The relation \eqref{randc} then follows by equating the residues in \eqref{YetaFundamentalrelationbis} for $\alpha_2=\frac{b}{2}$. The analyticity of $R(\alpha)$ in the region  $\alpha\in (\frac{b}{2},\frac{Q}{2})$ is then a consequence of the analyticity of $ C(\alpha,b, \alpha)$.

 \subsection{First shift equation for $R$}
It turns out that we can gain much more information from the strategy of studying the singularities of the left hand side of the identity  \eqref{YetaFundamentalrelation}. For this let us evaluate  \eqref{YetaFundamentalrelation}  at $\alpha_1=\frac{Q}{2}-\eta$ and $\alpha_3=\frac{1}{2b}+\eta$:  
\begin{equation}\label{CTused1}
C(\frac{1}{2b}-\eta,\alpha_2,\frac{1}{2b}+\eta)=T(\frac{Q}{2}-\eta,\alpha_2,\frac{1}{2b}+\eta)C(\frac{Q}{2}-\eta,\alpha_2+\frac{b}{2},\frac{1}{2b}+\eta)
\end{equation}
By the extended Seiberg bound \eqref{extseib} this makes sense  for fixed $\eta>0$ small enough and $\alpha_2\in(2\eta,\frac{1}{2b})$. As in the previous subsection   the tail analysis \eqref{theo4pointexpression1111} gives us a pole  \eqref{2pointfff}  of the structure constant when $\alpha_2\to 2\eta$. This comes from the contribution of the strongest singularity $\alpha_3=\tfrac{1}{2b}+\eta$  in the chaos integral in \eqref{cdefi}. When $\alpha_2 $ crosses this threshold another pole is produced in the same way by the next singularity for $\alpha_2 \to-2\eta$. Summarizing, the function
$$C(\frac{1}{2b}-\eta,\alpha_2,\frac{1}{2b}+\eta)-\frac{2R(\frac{1}{2b}+\eta)}{\alpha_2-2\eta}-\frac{2R(\frac{1}{2b}-\eta)}{\alpha_2+2\eta}$$ is analytic on a complex neighborhood of $(-2\eta-\delta, \frac{1}{2b})$ for some $\delta>0$ and coincides on $(2\eta,\frac{1}{2b})$ with
\begin{equation}\label{CTused2}
T(\frac{Q}{2}-\eta,\alpha_2,\frac{1}{2b}+\eta)C(\frac{Q}{2}-\eta,\alpha_2+\frac{b}{2},\frac{1}{2b}+\eta)-\frac{2R(\frac{1}{2b}+\eta)}{\alpha_2-2\eta}-\frac{2R(\frac{1}{2b}-\eta)}{\alpha_2+2\eta}
\end{equation}
 because of \eqref{CTused1}.
This latter function is thus analytic on $(-2\eta-\delta, \frac{1}{2b})$. By \eqref{2pointfff} the structure constant $C(\frac{Q}{2}-\eta,\alpha_2+\tfrac{b}{2},\frac{1}{2b}+\eta)$ appearing in expression \eqref{CTused2} has its first pole   appearing at $\alpha_2\to-2\eta$ and the pole is $\frac{2 R(\frac{Q}{2}-\eta)}{\alpha_2+2 \eta}$. Hence, multiplying  \eqref{CTused2} and taking the limit $\alpha_2\to-2\eta$ yields the relation
\begin{equation}\label{CTused3}
 T(\frac{Q}{2}-\eta,-2\eta,\frac{1}{2b}+\eta)R(\frac{Q}{2}-\eta)=R(\frac{1}{2b}-\eta).
\end{equation}
Some algebra about the function $T$ yields then the following shift equation: 
\begin{equation}\label{shift1}
-\mu\pi \frac{R(\alpha)}{l(-b^2)l(2b\alpha-b^2)l(2+2b^2-2b\alpha)}=R(\alpha-\frac{b}{2}).
\end{equation}
Since $R(\alpha)$ is analytic in $\alpha\in (\frac{b}{2},\frac{Q}{2})$ this identity holds in fact on  $\alpha\in (b,\frac{Q}{2})$ by uniqueness of analytic continuation and  then it allows us to prove that $R$ has a unique meromorphic extension to a complex neighborhood of the whole real line, which satisfies \eqref{shift1}.

\subsection{Analytic extension of the structure constant beyond the Seiberg bound}
An important consequence of the fusion relation \eqref{theo4pointexpression11one} with $\chi=\frac{b}{2}$ is that the ratio 
$$\frac{C(\alpha_1-\frac{b}{2},\alpha_2,\alpha_3)}{R(\alpha_1)C(Q-\alpha_1-\frac{b}{2},\alpha_2,\alpha_3)}$$ is equal to the right hand side of \eqref{Fundrelation} (where $A,B,C$ are given by \eqref{defabcfirst} with $\alpha_0=-\frac{b}{2}$) when $\alpha_1$ is smaller but close enough to $\tfrac{Q}{2}$ and $\alpha_1+\frac{b}{2}>\frac{Q}{2}$. Combining with the shift equation \eqref{shift1} for the reflection coefficient we obtain the relation (recall \eqref{Aformula})
\begin{equation}\label{reflect3point1}
R(\alpha_1+\frac{b}{2})C(Q-\alpha_1-\frac{b}{2},\alpha_2,\alpha_3)=-\frac{1}{\mu \pi}\mathcal{A}(b)C(\alpha_1-\frac{b}{2},\alpha_2,\alpha_3).
\end{equation}
This relation is reminiscent of \eqref{3pointconstanteqintro} valid for $\alpha_1+\frac{b}{2}<\frac{Q}{2}$. In both cases, the right-hand side is analytic in a neighborhood of $\alpha_1+\frac{b}{2}=\frac{Q}{2}$. Translating this analycity to the left-hand side implies  that the function
\begin{equation}\label{gluing}
\alpha_1\mapsto \left\{\begin{array}{ll} 
C(\alpha_1,\alpha_2,\alpha_3) & \text{if }\alpha_1<\frac{Q}{2}\\
R(\alpha_1)C(Q-\alpha_1,\alpha_2,\alpha_3) & \text{if }\alpha_1>\frac{Q}{2}
\end{array}\right.
\end{equation}
is the restriction of an analytic function to a neighborhood of $\alpha_1=\frac{Q}{2}$.

\subsection{Inversion relation}

Now we focus on the information we can get from  the crossing relation \eqref{Yeta2} in the region where it blows up. It was derived
for $\alpha_1,\alpha_2$ smaller but close to $\tfrac{Q}{2}$ and $\alpha_3=\tfrac{1}{2b}$. The left-hand side of \eqref{Yeta2} is analytic beyond the threshold $\alpha_1= \tfrac{Q}{2}$. The right-hand side has an analytic continuation in $\alpha_1$  given by \eqref{gluing}.
Invoking analytic continuation gives the relation
\begin{equation}
 C (\alpha_1-\frac{1}{2b}, \alpha_2, \frac{1}{2b})= \tilde T(\alpha_1,\alpha_2,\frac{1}{2b}) R(\alpha_1)R(\alpha_2) C(Q-\alpha_1,Q-\alpha_2 -\frac{1}{2b}, \frac{1}{2b}), 
\end{equation}
which is valid for $\alpha_1,\alpha_2$ close to $\tfrac{Q}{2}$ under the constraint that $\alpha_1,\alpha_2$ are respectively larger and smaller than $\tfrac{Q}{2}$.  Now we take the limit of both sides when $\alpha_2\to Q-\alpha_1$. Notice that along this asymptotic the expectation term $\E[ (\cdots)^{-s}  ]$ in the definition  \eqref{cdefi} of $C (\alpha_1-\frac{1}{2b}, \alpha_2, \frac{1}{2b})$ and $ C(Q-\alpha_1,Q-\alpha_2 -\frac{1}{2b}, \frac{1}{2b})$ converges to $1$ since in both cases sending  $\alpha_2\to Q-\alpha_1$ corresponds in the limit to summing the $3$ weights of insertions to $Q$, i.e. taking $s\to 0$. Hence the divergence comes from the $\Gamma$ function only and therefore 
\begin{equation}
\lim_{\alpha_2\to Q-\alpha_1}( Q-\alpha_1-\alpha_2)C (\alpha_1-\frac{1}{2b}, \alpha_2, \alpha_3)=-1,\quad \lim_{\alpha_2\to Q-\alpha_1}( Q-\alpha_1-\alpha_2)C(Q-\alpha_1,Q-\alpha_2 -\frac{1}{2b}, \alpha_3)=1.
\end{equation}
Combining with some elementary algebra  as $\alpha_2\to Q-\alpha_1$
$$( Q-\alpha_1-\alpha_2)l(\tilde{A})\to -b,\quad ( Q-\alpha_1-\alpha_2)^{-1}l(\tilde{A})\to \tfrac{1}{b}, \quad  l(\tilde{C})l(\tilde{A}+\tilde{B}-\tilde{C})\to l(\tilde{C})l(1-\tilde{C})=1$$
where $\tilde{A},\tilde{B},\tilde{C}$ are given by \eqref{defabc} with $\alpha_3=\frac{1}{2b}$. By definition \eqref{defT2} of $\tilde{T}$, this gives the inversion relation 
\begin{equation}\label{inversion}
 R(\alpha_1)R(Q-\alpha_1)=1
\end{equation}
for $\alpha_1$ in a neighborhood of $\frac{Q}{2}$, hence over a neighborhood of the real line by analyticity.

\subsection{Second shift equation}
Now we make use of the crossing relation \eqref{Yeta3} valid for $\alpha_1,\alpha_2\in (\tfrac{b}{2},\tfrac{Q}{2})$ and $\alpha_3\in (\tfrac{1}{2b},\tfrac{Q}{2})$. For  $\alpha_1$  close to $\tfrac{b}{2}$ then  $Q-\alpha_1 -\tfrac{1}{2b}$ is close to $\tfrac{Q}{2}$, namely the region where   the 3 point structure constant can be analytically extended with \eqref{gluing}. Hence,  analytic extension of \eqref{Yeta3}   below the threshold $\alpha_1=\tfrac{b}{2}$ produces the relation 
\begin{equation}
R(\alpha_1)R(Q-\alpha_1 -\tfrac{1}{2b}) C(\alpha_1 +\frac{1}{2b},\alpha_2, \alpha_3)= L( \alpha_1, \alpha_2, \alpha_3)R(\alpha_2) C(\alpha_1,Q-\alpha_2 -\frac{1}{2b}, \alpha_3)
\end{equation}
for $\alpha_1$ smaller but close to $\tfrac{b}{2}$. Using the inversion relation \eqref{inversion} $ R(Q-\alpha_1 -\frac{1}{2b}) =R(\alpha_1 +\frac{1}{2b})^{-1}$ we end up with
\begin{equation}\label{asymptotal}
R(\alpha_1)R(\alpha_1 +\frac{1}{2b})^{-1} C(\alpha_1 +\frac{1}{2b},\alpha_2, \alpha_3)= L( \alpha_1, \alpha_2, \alpha_3)R(\alpha_2) C(\alpha_1,Q-\alpha_2 -\frac{1}{2b}, \alpha_3).
\end{equation}
Now we evaluate this relation at $\alpha_3= \frac{Q}{2}-\alpha_1$ and want to take the limit $\alpha_2\downarrow \frac{b}{2}$ provided we compute the asymptotics of all terms involved in this relation. The asymptotics of $R(\alpha_2)$ can be obtained using the first shift equation \eqref{shift1} and  $l(2+b^2-2b \alpha_2)=-\frac{1}{2b(\alpha_2-\frac{b}{2})}(1+o(1))$
\begin{equation}\label{asymp1}
R(\alpha_2)=-\mu\pi \frac{R(\alpha_2+\frac{b}{2})}{l(-b^2)l(2b\alpha_2)l(2+ b^2-2b\alpha_2)}=\mu\pi \frac{R(b)}{2b\, l(-b^2)l(b^2)(\alpha_2-\frac{b}{2})}(1+o(1)).
\end{equation}
Once again, the limit
$$\lim_{\alpha_2\downarrow \frac{b}{2}}(\alpha_2-\frac{b}{2})C(\alpha_1 +\frac{1}{2b},\alpha_2, \tfrac{Q}{2}-\alpha_1)=1$$
comes from the divergence of the $\Gamma$ function in \eqref{cdefi} as $s=(\alpha_2-\frac{b}{2})/b$ converges to $0$ hence there is no contribution from the expectation term $\E[ (\cdots)^{-s}  ]$. The asymptotics of the 3 point structure constant $C(\alpha_1,Q-\alpha_2 -\frac{1}{2b}, \alpha_3)$ in the right-hand side of \eqref{asymptotal} is more involved as the  second insertion $Q-\alpha_2 -\frac{1}{2b}$ degenerates to the critical value $\frac{Q}{2}$. Yet the probabilistic representation makes the analysis possible and yields
\begin{equation}\label{structcomp}
 \lim_{\alpha_2\downarrow \frac{b}{2}}(\alpha_2-\tfrac{b}{2})C(\alpha_1,Q-\alpha_2 -\frac{1}{2b}, \tfrac{Q}{2}-\alpha_1)=-2. 
\end{equation}
Indeed analyzing  the probabilistic representation boils down to determining the asymptotics of the expectation $\E[ (\cdots)^{-s}  ]$  in  \eqref{cdefi} involving the integral \eqref{cdefi1}. The behaviour of this integral is completely dominated by a neighborhood of the second insertion with weight $Q-\alpha_2 -\frac{1}{2b}$ as it degenerates to the value  $\tfrac{Q}{2}$, hence requires to understand the behaviour of the quantity
$$\E \Big(\int_{|z-1|\leq \frac{1}{2}}\frac{1}{|z-1|^{2b(Q-2\epsilon)}}dM(z)\Big)^{\frac{\epsilon}{b}}$$
where we have set $\epsilon=\alpha_2-\frac{b}{2}$. The main idea is to perform a polar decomposition centered at $1$ of the GFF $\varphi$ (involved in the measure $M$). We will neglect the angular component as it does not contribute to the limit. Recalling  \eqref{circleaverage} and \eqref{Z1aa} we get the approximation (when neglecting the angular part of the GFF)
$$\E \Big(\int_{|z-1|\leq \frac{1}{2}}\frac{1}{|z-1|^{2b(Q-2\epsilon)}}dM(z)\Big)^{\frac{\epsilon}{b}}\approx \E \Big(\int_{ 0}^{\frac{1}{2}}\frac{1}{r^{2b(Q-2\epsilon)-1}}e^{2b\varphi_r(1)-2b^2\E\varphi_r(1)^2}\,dr\Big)^{\frac{\epsilon}{b}}.$$
It turns out that the process $r\mapsto \varphi_r(1)$ behaves like a Brownian motion in logarithmic time, namely like $r\mapsto B_{-\ln r}$.
Making the change of variables $u=-\ln  r$ we arrive at the expression
$$\E \Big(\int^{ \infty}_{\ln\frac{1}{2}} e^{2b(B_u-2\epsilon u)}\,du\Big)^{\frac{\epsilon}{b}}.$$
This type of integral is a standard object in  probability theory and is known to be dominated by the maximum of the Brownian path
$$\E \Big(\int^{ \infty}_{\ln\frac{1}{2}} e^{2b(B_u-2\epsilon u)}\,du\Big)^{\frac{\epsilon}{b}}\approx \E \Big( e^{2b\max_{u\geq 0}(B_u-2\epsilon u)}\Big)^{\frac{\epsilon}{b}}. $$
This maximum is exactly computable and has an exponential law with parameter $4\epsilon$. It is then a straightforward computation to obtain 
$$\E \Big(\int^{ \infty}_{\ln\frac{1}{2}} e^{2b(B_u-2\epsilon u)}\,du\Big)^{\frac{\epsilon}{b}}\to 2. $$
Combining with the other terms appearing in \eqref{cdefi}, it is then readily seen why the claim \eqref{structcomp} holds.

Finally some algebra gives
\begin{equation}\label{asymp3}
L( \alpha_1, \alpha_2, \frac{Q}{2}-\alpha_1)=-\frac{\alpha_2-\frac{b}{2}}{b}\frac{l(\frac{2\alpha_1-Q}{b})l(1+b^{-2})}{l( \frac{2\alpha_1}{b} )}(1+o(1)).
\end{equation}
Plugging the asymptotics \eqref{asymp1}+\eqref{structcomp}+\eqref{asymp3} into the relation \eqref{asymptotal} evaluated $\alpha_3= \frac{Q}{2}-\alpha_1$ establishes the relation
\begin{equation}\label{shift2}
R(\alpha_1)R(\alpha_1+\frac{1}{2b})^{-1}=-\mu\pi b^2 R(b)\frac{l(\frac{2\alpha_1-Q}{b})l(1+b^{-2})}{l( \frac{2\alpha_1}{b} )}=-\frac{c_b}{l( \frac{2\alpha_1}{b} )l(2+b^{-2}-\frac{2\alpha_1}{b})l(-b^{-2})}
\end{equation}
with $c_b=b^2\mu\pi R(b)$, where we have used $l(x)l(-x)=-x^{-2}$. This is our second shift relation for $R$.
 
\subsection{DOZZ formula for the reflection coefficient}
Now we can conclude our argument for the reflection coefficient. Set $\psi(\alpha)=\frac{R(\alpha)}{R_{{\rm DOZZ}}(\alpha)}$. This is a meromorphic function in a neighborhood of $\R$, which is $\frac{b}{2}$ periodic because both  $R(\alpha)$ and $R_{{\rm DOZZ}}(\alpha)$ satisfy the same shift equation \eqref{shift1}. Furthermore $\psi$ is positive on $\R$ because both  $R(\alpha)$ and $R_{{\rm DOZZ}}(\alpha)$ have same sign on $(\tfrac{b}{2},\tfrac{Q}{2})$ and by $\frac{b}{2}$-periodicity. The second shift equation entails that 
\begin{equation}\label{firstperiod}
\psi(\alpha)=C_b\psi(\alpha+\frac{1}{2b})
\end{equation}
 for some unknown positive constant $C_b$. Integrating both sides of \eqref{firstperiod} on an interval of size  $\frac{b}{2}$ and exploiting $\frac{b}{2}$-periodicity entails necessarily $C_b=1$. Then for incommensurable $\frac{b}{2}$, $\frac{1}{2b}$, which amounts to requiring $b^2\not\in\Q$, $\psi$ must be constant beacuse it has two incommensurable periods. The inversion relation \eqref{inversion} forces $\psi$ to be equal to $\pm 1$ and finally  $\psi=1$ because $\psi$ is positive. The case $b^2\in\Q$ follows by continuity in $b$.

\subsection{Proof of DOZZ formula}
Let us fix the values of the insertions $\alpha_2,\alpha_3$ very close to $\tfrac{Q}{2}$, say in the interval $(\tfrac{Q}{2}-\eta,\tfrac{Q}{2})$ for some small $\eta$. From the Seiberg bounds, the map $\alpha_1\in(2\eta,\tfrac{Q}{2})\mapsto F_{\alpha_2,\alpha_3} (\alpha_1):=C(\alpha_1,\alpha_2,\alpha_3)$ is analytic and, from the shift equation \eqref{3pointconstanteqintro}, it can be extended to a meromorphic function on the real line $\R$. Plugging the shift equation \eqref{shift1} for the reflection coefficient into    \eqref{theo4pointexpression11one} with $\chi=\frac{1}{2b}$ yields 
\begin{equation}\label{theomultiple}
\caT_{-\frac{1}{2b}
}(z)= C(\alpha_1-\frac{1}{2b}
 , \alpha_2,\alpha_3)  |F_{-}(z)|^2 +  \bar B(\alpha_1)R(\alpha_1+\frac{1}{2b})C(Q-\alpha_1- \frac{1}{2b}
,\alpha_2,\alpha_3) |\caF_{+}(z)|^2
\end{equation}
with (recall that $\tilde{\mu}=  \frac{(\mu \pi \ell(b^2)  )^{\frac{1}{b^2} }}{ \pi \ell(\frac{1}{b^2})}$)
\begin{align}\label{Bbardefin}
\bar B(\alpha)=-\tilde\mu   \frac{1}{  l(b^{-2}) l(2  \alpha_1/b)  l(2+b^{-2}- 2 \alpha_1/b) }.
 \end{align}
Identifying this relation with \eqref{Tsolution}, we get a relation between $C(\alpha_1-\frac{1}{2b} , \alpha_2,\alpha_3) $ and $ \bar B(\alpha_1)R(\alpha_1+\frac{1}{2b})C(Q-\alpha_1- \frac{1}{2b},\alpha_2,\alpha_3)$ as a consequence of \eqref{Fundrelation} and therefore a way to extend $ F_{\alpha_2,\alpha_3} $ beyond the value $\alpha_1=\frac{Q}{2}$. Yet this extension must satisfy \eqref{gluing} too. Uniqueness of analytic extension thus gives the second shift relation
\begin{align}\label{DOZZshift2}
C(\alpha_1+\frac{1}{2b},\alpha_2,\alpha_3)&=- \frac{_1}{^{\pi \tilde \mu}} \mathcal{A}(b^{-1})C(\alpha_1-\frac{1}{2b},\alpha_2,\alpha_3)
\end{align}
with $\mathcal{A}$ given by \eqref{Aformula}. 

The last argument is now to observe that the ratio $\alpha_1\mapsto \psi_{\alpha_2,\alpha_3} (\alpha_1):=\frac{ F_{\alpha_2,\alpha_3} (\alpha_1)}{C_{{\rm DOZZ}}(\alpha_1 , \alpha_2,\alpha_3)}$ is analytic on the real line as both functions involved in this ratio are analytic on $(\frac{Q}{2}-\eta,\frac{Q}{2})$ and satisfy the same shift equation \eqref{3pointconstanteqintro}, hence have the same set of simple poles/zeros. It is also $b$ and $1/b$ periodic because both functions satisfy the shift equations \eqref{3pointconstanteqintro}+\eqref{DOZZshift2}. In case these periods are incommensurable, meaning $b^2\not\in \Q$, $ \psi_{\alpha_2,\alpha_3}$ must be constant. Continuity in $b$   makes sure that $ \psi_{\alpha_2,\alpha_3}$ is actually constant for all values of $b$. Furthermore, symmetry of $\psi$ in its arguments $\alpha_i$'s forces $\psi$ to be constant, eventually only depending on $b$. The value of the constant may be identified to be $1$ as both functions $C_{{\rm DOZZ}}(\alpha_1 , \alpha_2,\alpha_3)$ and $C(\alpha_1 , \alpha_2,\alpha_3)$ satisfy \eqref{2pointf}.

\section{Fusion with degenerate field}\label{asyana}
In this section we will outline the argument leading to Theorem \ref{thetheo4pointexpression11one} in the case $\chi=\frac{b}{2}$. The case $\chi=\frac{1}{2b}$ is similar. 
As explained in Section \ref{bpzsec}, $\caT_{-\frac{b}{2}}$ is of the form \eqref{Tsolution}. We determine the coefficients $\lambda_i$ by studying the  small $z$ behaviour of $\caT_{-\frac{b}{2}}(z)$ given by \eqref{Tdefi} with $s=\frac{\sum_{i=1}^3 \alpha_i-Q}{b}-\frac{1}{2}$ (recall that $\alpha_0=-\frac{b}{2}$ in this context). All the weights are assumed to satisfy the Seiberg bounds and we suppose $\alpha_1+\frac{b}{2}>\frac{Q}{2}$. For simplicity we also assume  $\alpha_1$ is close enough to $\frac{Q}{2}$  so that  $2b (Q-2\alpha_1)\in (0,1)$. We will then show
\begin{equation}\label{tresult}
\mathcal{T}(z) = C(\alpha_1-\frac{b}{2}, \alpha_2,\alpha_3)+ R(\alpha_1)  |z|^{2b(Q-2\alpha_1)} C(Q-\alpha_1-\frac{b}{2},\alpha_2,\alpha_3)+o(  |z|^{2b(Q-2\alpha_1)} ).
\end{equation}

It is first instructive to contrast the case $\alpha_1+\frac{b}{2}>\frac{Q}{2}$ to the simpler case $\alpha_1+\frac{b}{2}<\frac{Q}{2}$ corresponding to the fusion rule \eqref{fusion} without reflection. In both cases we need to expand 
$\E  [ r(z)^{-s}]$ in \eqref{Tdefi} around $z=0$. Write $r(z)=r(0)+(r(z)-r(0))$. Then to first order we have
\begin{align}\nonumber
\E [ r(z)^{-s}]-\E [r(0)^{-s}]&\approx s\E [ (r(0)-r(z)) r(0)^{-s-1}]\\&=s \int_\C  \frac{ |x|^{2b^2}- |x-z|^{2b^2}}{ |x|^{4b \alpha_1} |x-1|^{4b\alpha_2}  }  g(x)^{1-
b \sum_{i=0}^3 \alpha_i } \E\, [e^{2b\varphi(x)- 2b^2 \E[\varphi(x)^2] }
r(0)^{-s-1} ]
dx\label{1stord}
\end{align}
where $\alpha_0=-\frac{b}{2}$ and for clarity of exposition we use the heuristic expression $e^{2b\varphi(x)- 2b^2 \E[\varphi(x)^2] }dx$ for the measure $dM(x)$. 
The last expectation can be evaluated by the complete the square trick (or Girsanov in probability theory) as before namely by a shift in the Gaussian field $\varphi(y) \rightarrow \varphi(y)+ 2bG(y,x)$ (recall that $G(x,y)= \ln \frac{1}{|x-y|}-\frac{1}{4} (\ln g(x)+\ln g(y))$) :
\begin{align*}
 \E \, [ e^{2b \varphi(x)- {2b^2} \E[\varphi(x)^2] }
r(0)^{-s-1} ]
&
= \E  \left [ \left  (\int_{\C}  \frac{g(x)^{-b^2}}{ |y|^{4b(\alpha_1-\frac{b}{2})} |x-y|^{4b^2}|y-1|^{4b \alpha_2}  }g(y)^{1
- b \sum_{l=1}^3\tilde \alpha_l }  e^{2b \varphi(y)-2b^2 \E[\varphi(y)^2] }
dy   \right )^{-\tilde s}  \right ]
\end{align*}
where $\tilde \alpha_1=\alpha_1+\frac{b}{2}$, $\tilde \alpha_i=\alpha_i$ for $i=2,3$  and $\tilde s=(\sum_{i=1}^3\tilde\alpha_i-Q)/b$. Now make a change of variables $x=zu$ in the above relation. This results in the following expansion around $z=0$ ($|zu-y|$ goes to $0$ for $u$ and $y$ fixed)
\begin{align}\label{1stord1}
\E [ r(z)^{-s}]-\E [r(0)^{-s}]&= s|z|^{2b(Q-2\alpha_1)}\int_\C  \frac{ |u|^{2b^2}- |u-1|^{2b^2}}{ |u|^{2b \alpha_1} }du 
\\&\times
 \E  \left [ \left (\int  \frac{1}
 { |y|^{2b (\alpha_1+\frac{b}{2})} |y-1|^{2b\alpha_2}  } g(y)^{1
- b\sum_{l=1}^3\tilde \alpha_l }  e^{2b \varphi(y)- 2b^2\E[\varphi(y)^2] }
dy  \right )^{-\tilde s}  \right ]+o(|z|^{2b(Q-2\alpha_1)})\nonumber.
\end{align}
For   $\alpha_1+\frac{b}{2}<\frac{Q}{2}$, the last expectation is the one occurring in $C(\alpha_1+\frac{b}{2},\alpha_2,\alpha_3)$ and using \eqref{formuleint2} some algebra then yields
 \begin{equation}\label{theo4pointexpression11}
\caT(z)= C(\alpha_1-\frac{b}{2}, \alpha_2,\alpha_3) + |z|^{2b(Q-2\alpha_1)}B(\alpha_1)C(\alpha_1+\frac{b}{2},\alpha_2,\alpha_3)+o(|z|^{2b(Q-2\alpha_1)})
\end{equation}
consistent with  the fusion rule \eqref{fusion}. Note that this argument breaks down when $\alpha_1+\frac{b}{2}>\frac{Q}{2}$. Indeed,  then $2b\alpha_1>2$ and the integral in \eqref{1stord1} diverges around $u=0$ in that case. Thus the perturbative expansion \eqref{1stord} is no more valid and we need a more sophisticated argument for the small $z$ behaviour of $\E\, [ r(z)^{-s}]$. 

We proceed by  splitting  $r(z)=r_1(z)+r_2(z)$ where 
\begin{equation*}
r_1(z)=  \int_{|x| > |z|} e^{2b \varphi(x)-2b^2 \E[\varphi(x)^2] } \frac{|x-z|^{2b^2}}{ |x|^{4b \alpha_1}  |x-1|^{4b \alpha_2}  }  {g}(x)^{1
- b \sum_{i=0}^3 \alpha_i } d^2x.
\end{equation*}
and 
\begin{equation*}
r_2(z)=  \int_{|x| < |z|} e^{2b \varphi(x)-2b^2 \E[\varphi(x)^2] } \frac{|x-z|^{2b^2}}{ |x|^{4b \alpha_1}  |x-1|^{4b \alpha_2}  }  {g}(x)^{1
- b \sum_{i=0}^3 \alpha_i } d^2x.
\end{equation*}
Then write
\begin{align}
\E [ r(z)^{-s}]-\E [r(0)^{-s}]= \E [(r_1(z)+r_2(z))^{-s}-r_1(z)^{-s}]+ \E (r_1(z)^{-s}-r_1(0)^{-s}]\label{1stordd}
\end{align}
where we noted that  $\E [ r(0)^{-s}]=\E [r_1(0)^{-s}]$. The first expectation on the RHS turns out to be the leading one and for it we need to  understand the tail behaviour of the random variable $r_2(z)$ as $z\to 0$. Recall the definition \eqref{circleaverage} of 
 the average of $\varphi$ on a circle of center $z$ and radius $\epsilon$. Then $\varphi_{|z|}(0)$ and $\varphi(x)-\varphi_{|z|}(0)$ are independent for $|x| \leq |z|$ as can be seen on a simple computation on covariance of the field:
\begin{equation}\label{circleav}
\E [ \varphi_{|z|}(0)(\varphi(x)-\varphi_{|z|}(0))]=0.
\end{equation}
We then get (using $g(0)=1$) around $z=0$
\begin{align}\label{r2app}
r_2(z)  & \approx    e^{2b\varphi_{|z|}(0)-2b^2\E[  \varphi_{|z|}(0)^2    ] }J_{\alpha_1}\\
J_{\alpha_1}&=  \int_{|x| < |z|} e^{2b(\varphi(x)-\varphi_{|z|}(0))    -2b^2 \E[(\varphi(x)-\varphi_{|z|}(0))^2] } \frac{ |x-z|^{2b^2}}{ |x|^{4b \alpha_1 }}d^2x \nonumber
\end{align}
where "$\approx$" from now on means that we ignore corrections to the leading small $z$ asymptotics.  The covariance of the field $\varphi(|z|x)-\varphi_{|z|}(0)$ equals $\ln|x-y|^{-1}$ for $|x| \leq 1$. 
By a change of variables $x=|z|u$ we get in distribution   
\begin{align*}
J_{\alpha_1} =  |z|^{2b(Q -2\alpha_1)} 
   \int_{|u| < 1} e^{2b (\varphi(|z|u)-\varphi_{|z|}(0))   - 2b^2 \E [ (\varphi(|z|u)-\varphi_{|z|}(0))^2] }  \frac{|u-1|^{2b^2}}{ |u|^{4b\alpha_1}   } d^2u:= |z|^{2b(Q -2\alpha_1)}
   I_{\alpha_1}.
 \end{align*}
 and thus
  \begin{align}\label{r2app}
r_2(z)  \approx  |z|^{2b(Q -2\alpha_1)}
 e^{2b\varphi_{|z|}(0)-2b^2 \E[  \varphi_{|z|}(0)^2    ] } I_{\alpha_1}:=\epsilon  I_{\alpha_1}.
\end{align}
with $\epsilon=  |z|^{2b(Q -2\alpha_1)}
 e^{2b\varphi_{|z|}(0)-2b^2 \E[  \varphi_{|z|}(0)^2    ] }$. The distribution of the random variable $ I_{\alpha_1}$ is independent of $z$ and has the same tail as $Z(\alpha_1)$ in \eqref{theo4pointexpression1111} (because the tail of $ I_{\alpha_1}$ is concentrated around $u=0$ in the integral $\int_{|u|<1} \cdots$ defining $I_{\alpha_1}$) i.e. its PDF $p_{\alpha_1}$  is given by
\begin{equation}\label{tail}
p_{\alpha_1}(u)  \underset{u \to \infty}{\sim} \tfrac{(Q-2\alpha_1)}{b} \bar R(\alpha_1) \frac{1}{  u^{\frac{(Q-2\alpha_1)}{b}+1}} .
\end{equation}
 Finally, we may assume in \eqref{r2app} that $I_{\alpha_1}$ is independent of both  $r_1(z)$ and $\varphi_{|z|}(0)$. The reason we can assume this is that we can restrict the integral which defines $r_1(z)$ to $|x| > |z|^\beta$ with $\beta<1$ and for such $x$ and $|u| \leq 1$ the covariance $\E[  (\varphi(|z|u)-\varphi_{|z|}(0)) \varphi(x)   ]$ goes to $0$ for $|z| \to 0$. With these approximations we then get
\begin{align}\nonumber
 \E [ (r_1(z)+r_2(z))^{-s}-r_1(z)^{-s}]&= \E  \left [ \int_0^\infty p_{\alpha_1}(u) ((r_1(z)+\epsilon u)^{-s}-r_1(z)^{-s})du  \right ]\\ 
&
\approx \frac{(Q-2\alpha_1)}{b} \bar R(\alpha_1) \int_0^\infty \left (  \frac{1}{(1+v )^s}-1  \right ) \frac{1}{  v^{\frac{Q-2\alpha_1}{b}+1}} dv\, \E\, [ r_1(z)^{-s} (\frac{\epsilon}{r_1(z)} )^{\frac{Q-2\alpha_1}{b}}] \nonumber
\\ 
&  =  \frac{(Q-2\alpha_1)}{b} \bar R(\alpha_1)\Gamma(\frac{2\alpha_1-Q}{b})\frac{  \Gamma (s+\frac{Q-2\alpha_1}{b})}{   \Gamma (s) }
 \E\, [ {\epsilon}^{\frac{(Q-2\alpha_1)}{b}}  r_1(z)^{-s-\frac{(Q-2\alpha_1)}{b}}] .\label{approx}
\end{align}
where we have computed the $dv$ integral with \eqref{gamma1}. Using $ \E[  \varphi_{|z|}(0)^2]=  -\ln|z|$ we get
\begin{align*}
 {\epsilon}^{\frac{Q-2\alpha_1}{b}} &=|z|^{2(Q-2\alpha_1)^2} (e^{2b \varphi_{|z|}(0)-2b^2 \E  \varphi_{|z|}(0)^2     })^\frac{Q-2\alpha_1}{b}\\
 &=   |z|^{2b(Q-2\alpha_1)}e^{2(Q-2\alpha_1) \varphi_{|z|}(0)-2(Q-2\alpha_1)^2 \E  \varphi_{|z|}(0)^2     }
\end{align*}
Thus denoting $\hat s=s+\frac{Q-2\alpha_1}{b}=(-\alpha_1-\frac{b}{2}+\alpha_2+\alpha_3)/b$  we need to compute the expectation
\begin{align*}
  \E \frac{1}{r_1(z)^{\hat s}}  (e^{2(Q-2\alpha_1) \varphi_{|z|}(0)-2(Q-2\alpha_1)^2 \E[  \varphi_{|z|}(0)^2    ] })\approx    \E \frac{1}{r_3(z)^{\hat s}} 
\end{align*}
where we shifted again the field in the definition of $r_3(z)$ (by the Girsanov theorem)
\begin{equation*}
r_3(z)=  \int_{|x| > |z|} e^{2b \varphi(x)- 2b^2 \E[\varphi(x)^2] } \frac{|x-z|^{2b^2}e^{4b(Q-2\alpha_1)G(z,x)}}{ |x|^{4b \alpha_1}  |x-1|^{4b\alpha_2}  }  {g}(x)^{1- b \sum_{i=0}^3 \alpha_l } dx.
\end{equation*}

Using the fact that $G(z,x)= \ln \frac{1}{|z-x|}-\frac{1}{4}  (\ln g(z)+\ln g(x))$ we get around $z=0$
\begin{align*}
r_3(z)&\approx
 \int_\C e^{2b \varphi(x)-2b^2\E[\varphi(x)^2] } \frac{1}{ |x|^{4b (Q-\alpha_1-\frac{b}{2})} |x-1|^{4b\alpha_2}  }  {g}(x)^{1 - b (Q-\alpha_1-\frac{b}{2}+\alpha_2+\alpha_3) } dx\\&
 =
 \rho(Q-\alpha_1-\tfrac{b}{2},\alpha_2,\alpha_3).
\end{align*}
Combining these calculations  and recalling the definition of the reflection coefficient \eqref{deffullR} we arrive at 
\begin{align}\nonumber
 \E [ (r_1(z)+r_2(z))^{-s}-r_1(z)^{-s}]&= \mu^{\frac{2\alpha_1-Q}{b}}R(\alpha_1)\frac{ \Gamma (\hat s)}{   \Gamma (s) }
  \E\, [\rho(Q-\alpha_1-\tfrac{b}{2},\alpha_2,\alpha_3)^{-\hat s}] |z|^{2b(Q-2\alpha_1)}\\&+o( |z|^{2b(Q-2\alpha_1)}).
  \label{approx}
\end{align}
Let us next consider the second difference in  the RHS of \eqref{1stordd}. We can repeat the calculation we did in the case $\alpha_1+\frac{b}{2}<\frac{Q}{2}$ with the difference that \eqref{1stord1} is replaced by
\begin{align}\nonumber
\E [ r_1(z)^{-s}]-\E [r_1(0)^{-s}]&\approx s|z|^{2b(Q-2\alpha_1)}\int_\C  \frac{ |u|^{2b^2}- |u-1|^{2b^2}1_{|u|>|1|}}{ |u|^{4b \alpha_1} }du 
\\&
 \E \left [ \left (\int_\C  \frac{1}
 { |y|^{4b \alpha_1} |y-z|^{2b^2} |y-1|^{4b \alpha_2}  } {g}(y)^{1
- b \sum_{l=1}^3\tilde \alpha_l }  e^{2b\varphi(y)- 2b^2 \E[\varphi(y)^2] }
dy \right )^{-\tilde s} \right ]=o(|z|^{2b(Q-2\alpha_1)})\label{1stord2}
\end{align}
since the expectation tends to zero as $z\to 0$ and now the $u$-integral converges due to the cutoff $1_{|u|>|1|}$. Combining \eqref{approx} and \eqref{1stord2} with 
\eqref{Tdefi} we get
\begin{align*}
  \mathcal{T}_{-\frac{b}{2}}(z) - \mathcal{T}_{-\frac{b}{2}}(0)&=
b^{-1} \mu^{-\tilde s} 
R(\alpha_1) \Gamma (\tilde s)
  \E\, \rho(Q-\alpha_1-\tfrac{b}{2},\alpha_2,\alpha_3)^{-\tilde s} |z|^{2b(Q-2\alpha_1)}+o( |z|^{2b(Q-2\alpha_1)})\\&=R(\alpha_1)  C(Q-\alpha_1-\frac{b}{2},\alpha_2,\alpha_3)|z|^{2b(Q-2\alpha_1)}+o( |z|^{2b(Q-2\alpha_1)})
 \end{align*}
yielding \eqref{tresult}.

\appendix
\section{Some integral relations}
In this section we recall  the relation for $\alpha,\beta>0$ and $1<\alpha+\beta<3/2$
\begin{equation}\label{formuleint2}
\int_{\R^2}  |z|^{2(\alpha-1)} \big(|z-1|^{2(\beta-1)}-|z|^{2(\beta-1)}   \big)   dz  = \pi \frac{1}{l(1-\alpha) l(1-\beta)  l(\alpha+\beta) }
\end{equation} 

 and for all $p>0$ and $a \in (1,2)$ the relation
 
\begin{equation}\label{gamma1}
 \int_0^{\infty}    \left (  \frac{1}{(1+v )^p}-1  \right ) \frac{1}{v^a}  dv=   \frac{\Gamma(-a+1)  \Gamma (p+a-1)}{   \Gamma (p) }
\end{equation}

%
%

%
%

\section{Probabilistic representation of the reflection coefficient}\label{PRR}

The tail coefficient $ \bar R(\alpha)$ appearing in \eqref{theo4pointexpression1111} is given in terms of multiplicative chaos
\begin{equation}\label{defunitR}
 \bar{R}(\alpha)=  \E \left [ \left ( \int_{-\infty}^\infty  e^{   2b \mathcal{B}_s^\alpha  } Z_s ds  \right )^{\frac{Q-2\alpha}{b}} \right].
\end{equation}
where $\mathcal{B}_s$ is a two sided Brownian motion with negative drift $2\alpha-Q$ conditioned to stay negative, i.e. 
\begin{equation}\label{BMneg}
 \mathcal{B}^\alpha_s = \left\{
 \begin{array}{ll}
  B^\alpha_{-s} & \text{if } s < 0\\
    \bar{B}^\alpha_{s} & \text{if } s >0 \end{array} \right.
\end{equation}
where $(B^{\alpha}_s)_{s \geq 0},(\bar B^{\alpha}_s)_{s \geq 0}$ are  two independent standard Brownian motions starting from $0$ with negative drift $2\alpha-Q$ and conditioned to stay negative. The process $Z_s$ is an independent stationary process formally given by
$$
Z_s=\int_0^{2\pi}e^{2bY(s,\theta)-2b^2\E Y(s,\theta)^2}d\theta
$$
whre $Y(s,\theta)$ is a Gaussian field with covariance
\begin{equation}\label{covlateral}
\E[  Y(s,\theta) Y(t,\theta') ]  = \ln \frac{e^{-s}\vee e^{-t}}{|e^{-s}e^{i \theta} - e^{-t} e^{i \theta'} |}.
\end{equation}

\end{document}